\documentclass[ reprint, superscriptaddress,  amsmath,amssymb,  aip,  apl
]{revtex4-1}%
\usepackage{graphicx}
\usepackage{bm}
\usepackage{hyperref}
\usepackage{amsmath}
\usepackage{amsfonts}
\usepackage{amssymb}%

\begin{document}

\title{Broadband polarization transformation via enhanced asymmetric
transmission through arrays of twisted complementary split-ring resonators}
\author{Zeyong Wei}
\affiliation{Tongji University,Shanghai,200092,China}
\affiliation{Key Laboratory of Advanced Micro-structure Materials(MOE) and Department of
Physics, Tongji University, Shanghai 200092, China}
\author{Yang Cao}
\affiliation{Tongji University,Shanghai,200092,China}
\affiliation{Key Laboratory of Advanced Micro-structure Materials(MOE) and Department of
Physics, Tongji University, Shanghai 200092, China}
\author{Yuancheng Fan}
\affiliation{Tongji University,Shanghai,200092,China}
\affiliation{Key Laboratory of Advanced Micro-structure Materials(MOE) and Department of
Physics, Tongji University, Shanghai 200092, China}
\author{Xing Yu}
\affiliation{Tongji University,Shanghai,200092,China}
\affiliation{Key Laboratory of Advanced Micro-structure Materials(MOE) and Department of
Physics, Tongji University, Shanghai 200092, China}
\author{Hongqiang Li}
\email{hqlee@tongji.edu.cn}
\affiliation{Tongji University,Shanghai,200092,China}
\affiliation{Key Laboratory of Advanced Micro-structure Materials(MOE) and Department of
Physics, Tongji University, Shanghai 200092, China}

\begin{abstract}
This study proposes an ultrathin chiral metamaterial slab stacked with twisted
complementary split-ring resonators (CSRRs) for highly efficient broadband
polarization transformation. The polarization of linearly polarized
electromagnetic waves can be rotated in a specific direction by passing it
through such a slab having a thickness of about one-tenth the operational
wavelength. Microwave experiments verified the theoretically predicted
conversion efficiency of up to 96{\%} covering a bandwidth of 24{\%} of the
central wavelength. CSRRs with circular symmetry provide increased interlayer
coupling strength, which produces a high-efficiency broadband response and
strong isolation of the original polarization.
\end{abstract}
\maketitle

Metamaterials \cite{1} are potentially useful for tuning the polarization of
electromagnetic (EM) waves because they offer myriad ingredients and options
beyond conventional EM materials. Two mechanisms in particular, which use
either the birefringence effect of anisotropic metamaterials
\cite{2,3,4,5,6,7,8,9,10} or the optical activity of chiral metamaterials
\cite{11,12,13,14,15,16,17}, can be exploited in the metamaterial regime.
More interestingly, an ultrathin metamaterial slab can be used for
polarization transformation and perfectly asymmetric transmission without
transmission attenuation \cite{9}. However, in previous studies, the novel
functionality of ultrathin slabs operated only at a single frequency
\cite{3,4,7,8,9,11,12,13,14,15,16,18}. Extension of the operational bandwidth
of such a thin polarization transformer is desirable.

A strategy of stacking metallo-dielectric multilayers has been shown to
dramatically extend the transmission peak to a broad transparency band when
the guided resonance modes of drilled apertures on the metallic layers
dominate the enhanced transmission through the multilayered system \cite{19}.
Here, a similar scheme consisting of multilayered system containing arrays of
twisted complementary split-ring resonators (CSRRs) is proposed for broadband
polarization transformation (BPT).

The structure we investigated is comprised of three metallic layers and two
dielectric spacer layers. Figures 1(a) and 1(b) present a front-view photo of
our sample and the schematic of a unit cell, respectively. The metallic layers
lie in the $xy$ plane; each has a thickness of $t=0.035\text{mm}$ and is
perforated by an array of CSRRs. The dielectric layers each have a thickness
of $h=1.55\text{mm}$ and a dielectric constant of $\varepsilon_{r} =2.65$. A
CSRR unit is constructed by etching two concentric semi-annular rings through
the metallic layer; Figures 1(c)$\sim$1(e) present front views of the bottom,
middle, and top metallic layers, respectively. The rings in a unit are
symmetrically segmented by two metallic strips with a width of $g=0.2\text{mm}%
$. The CSRRs have a lattice constant of $p=10\text{mm}$ and inner and outer
radii of $r=3.8\text{mm}$ and $R=4.8\text{mm}$, respectively. A unit cell
contains three CSRRs aligned along the $z$ axis with zero displacement in
the$xy$ plane and arranged from the bottom layer to the top layer with
anticlockwise twist angles of $\alpha=0,\theta/2\text{ and }\theta(\theta$ =
90\r{ } for our sample).

\begin{figure}[pb]
\begin{center}
\includegraphics[
width=6.5cm
]{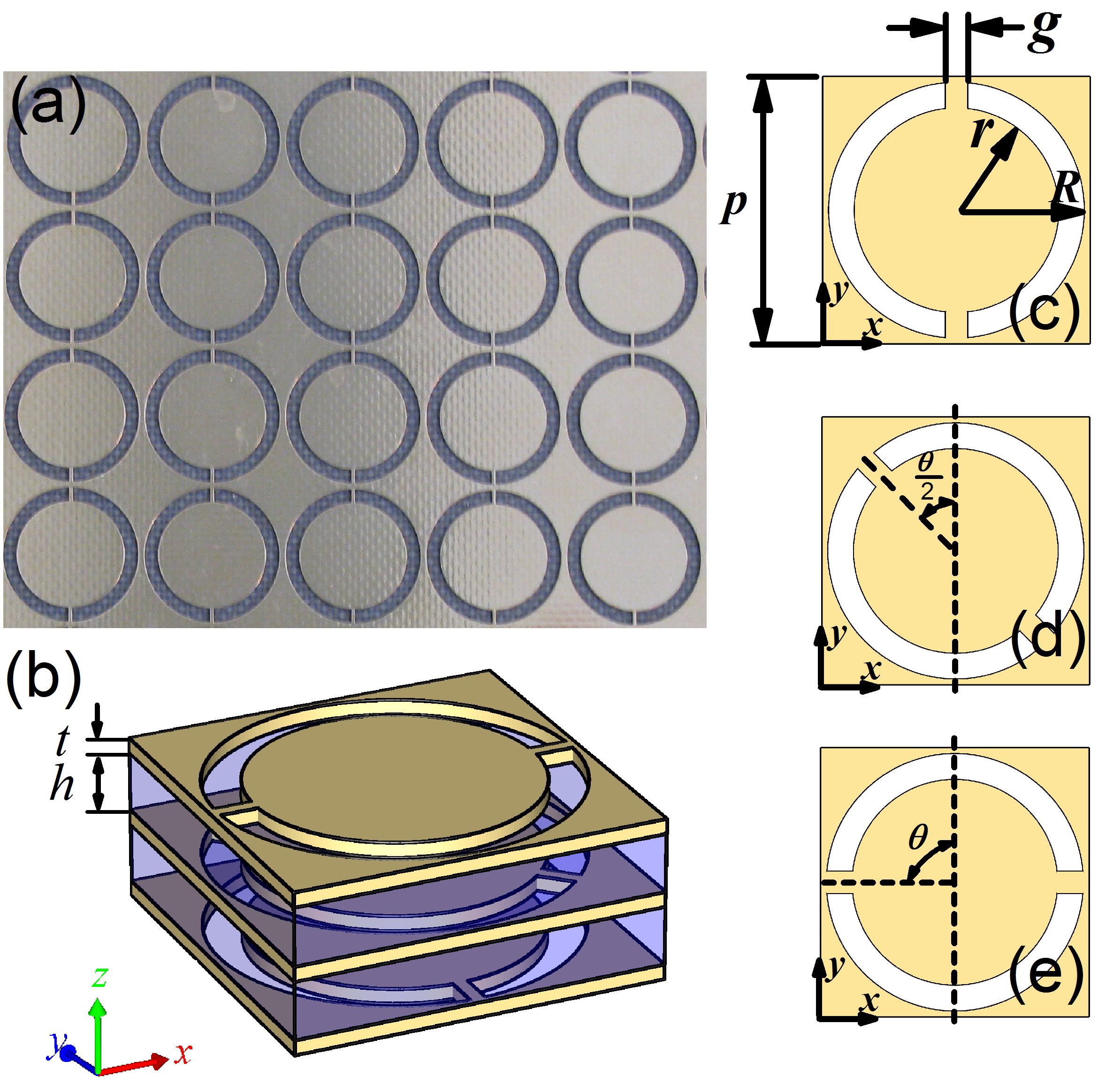}
\end{center}
\caption{(a) Front-view photo and (b) 3D schematic of a unit cell of the
stacked CSRR system. (c)$\sim$(e) Front view of three CSRR units with twist
angles of $0,\theta /2\text{ and }\theta $, respectively, at the bottom,
middle, and top metallic layers. Yellow indicates metallic regions.}%
\end{figure}

Assuming the metals are perfect electric conductors, we can employ the modal
expansion method (MEM) \cite{20,21,22,23,24,25} to semi-analytically solve the
transmission and reflection spectra and band structures of the multilayered
system. The EM fields within the metallic layers exist only inside the
symmetric CSRRs, and the in-plane field components can be rigorously expressed
as a series of eigenmodes of the ring apertures. Since the width $g$ of the
metallic strips is much smaller than both the outer and inner radii ($g\ll
r,g\ll R)$, the eigenmodes inside the CSRRs can be approximated by the
transverse electric eigenmodes $\text{TE}_{mn}(m,n=1,2,...)$ of a conventional
ring aperture as integer orders of Bessel functions \cite{26}. The
calculations converge quickly provided that only three lowest eigenmode
orders($\text{TE}_{11}$, $\text{TE}_{21}$ and $\text{TE}_{31})$ are adopted.

EM waves normally incident on the bottom metallic layer, propagating along the
$z^{+}$ direction, are $x$-polarized along the orientation of the CSRRs (the
line across a CSRR's metallic strips) in the incident plane. The intensity of
$x$- or $y$-polarized transmitted waves is defined as $T_{ix} =\vert
E_{i}^{Tran} /E_{x}^{Inc} \vert^{2}(i=x,y)$, where $E_{x}^{Inc} $ is the
electric field of the incident wave, and $E_{i}^{Tran} (i=x,y)$ is the $x$ or
$y$ component of the electric field of the transmitted waves.

Figure 2(a) presents the calculated and measured transmission spectra. The
calculated intensity of $x$-polarized transmitted waves $T_{xx} $ (red dashed
line) is less than $0.6\% $ at all frequencies, whereas that of $y$-polarized
transmitted waves $T_{yx} $ (blue solid line) is greater than $96\% $ within a
frequency range of 9.8$\sim$12.5 GHz. That is, within this frequency range,
the $x$-polarized incident waves are perfectly transformed to $y$-polarized
transmitted waves. The bandwidth is about 24{\%} of the central frequency.
Within this band, a system having a thickness of about one-tenth the
operational wavelength is nearly reflectionless.

\begin{figure}[ptb]
\begin{center}
\includegraphics[
width=6.5cm
]{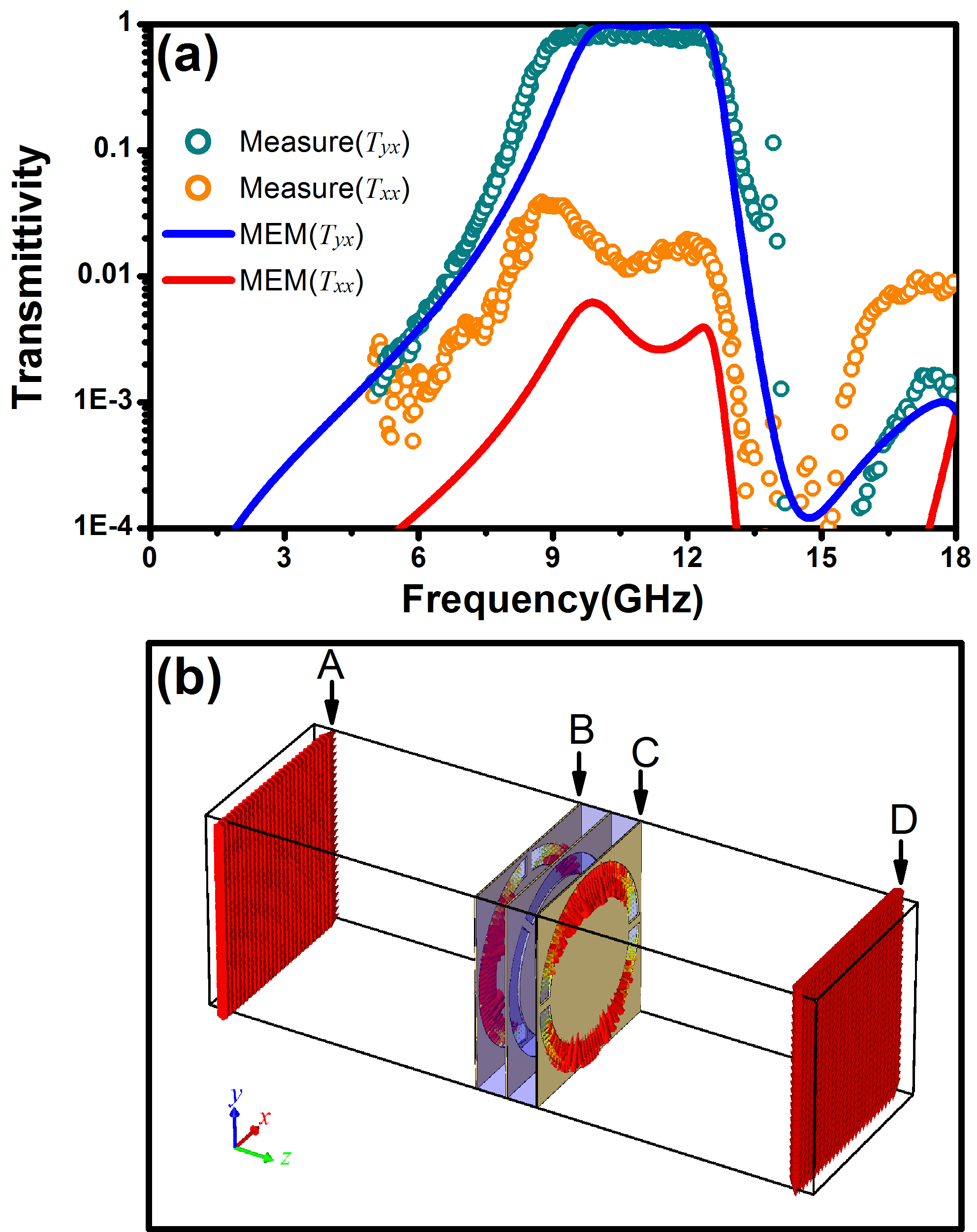}
\end{center}
\caption{(a) Calculated and measured transmission spectra of $T_{yx} $ and
$T_{xx} $. (b) Calculated spatial distribution of electric fields for 11-GHz
$x$-polarized EM wave incident along the $z^+$ direction from the left side
(bottom surface) of the unit-cell. Slices A and D are each 15 mm from the
bottom and top sample surfaces, while Slices B and C are set at the surfaces
of the first and third metallic layers.}%
\end{figure}

For verification, we fabricated a $250\times250\text{mm}^{2}$ sample on a
printed circuit board with the same structural parameters as the theoretical
model. The transmission spectra, taken through the sample plate and measured
by a vector network analyzer (Agilent 8722ES), were normalized using the
transmission coefficients in free space. The measured results [circles and
squares for $T_{yx}$ and $T_{xx}$, respectively, in Fig. 2(a)] verified the
theoretical predictions quite well. The evolution of EM transportation inside
the structure was visualized by calculating the spatial field distributions at
11 GHz [Fig. 2(b)]. An $x$-polarized wave incident from the left (Slice A)
excites the guided resonant modes of the corresponding CSRRs of the bottom
(Slice B), middle, and top (Slice C) metallic layers and finally is perfectly
transformed to a $y$-polarized transmitted wave (Slice D). The field
distributions in each layer explicitly indicate that the excitation of local
resonant modes and interlayer coupling among the twisted CSRRs are crucial to
the BPT functionality.

\begin{figure}[ptb]
\begin{center}
\includegraphics[
width=6.5cm
]{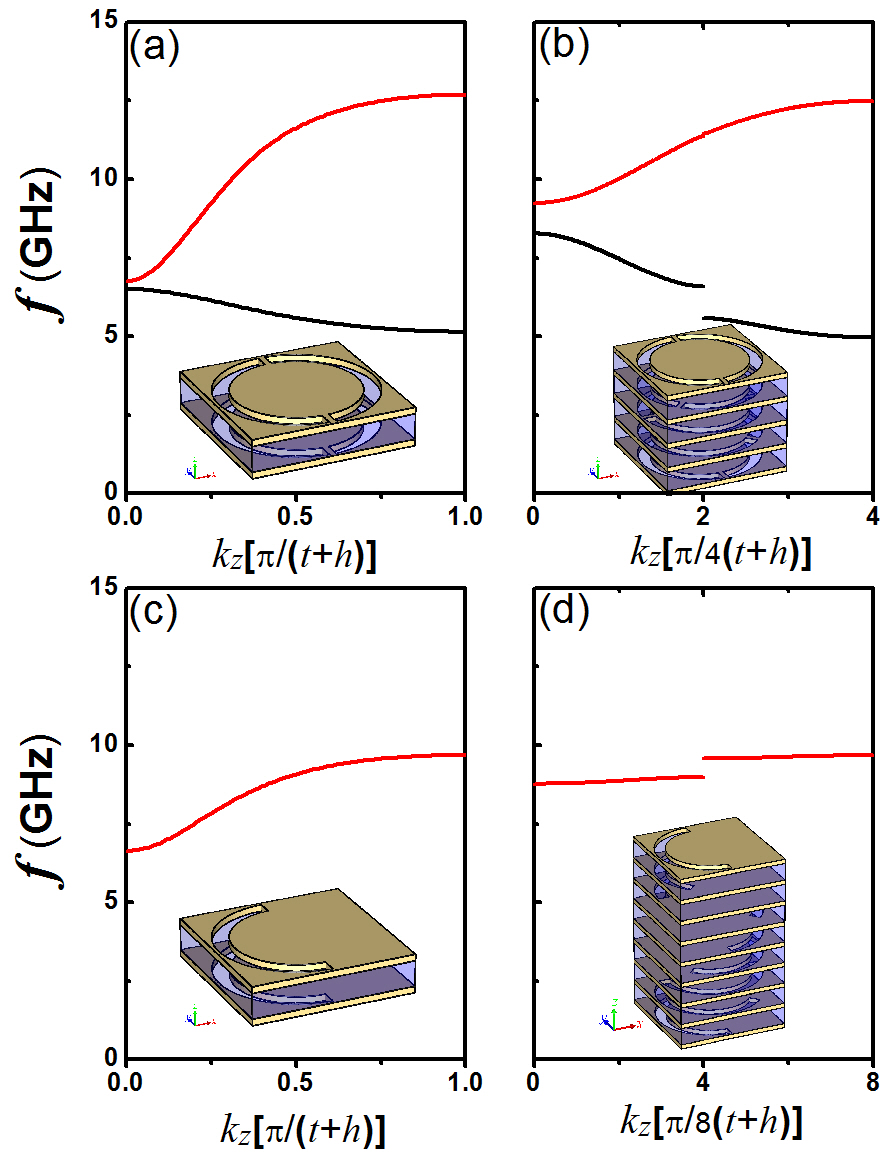}
\end{center}
\caption{Dispersions along the $z$ direction for the CSRR periodic systems shown
in the insets. (a) $\theta =0$, (b) $\theta =90^\circ $ with two-fold
circular symmetry, (c) $\theta =0$, (d) $\theta =90^\circ $with only one
CSRR unit at each metallic layer.}%
\end{figure}

The formation of the broad pass-band (9.8$\sim$12.5 GHz) of $T_{yx}$ can be
analyzed heuristically by calculating the band structure of the multilayered
CSRR system. The $T_{yx}$ pass-band in Fig. 2(a) precisely falls inside the
frequency range with the highest branch in Fig. 3(b) which shows the band
structures corresponding to the periodic model [inset, Fig. 3(b)] derived from
our twisted model system. In contrast, the periodic model with zero twist
angle ($\theta=0^{\text{o}})$, as shown in Fig. 2(a), exhibits a very broad
pass-band, as discussed in a previous work [19]. The results shown in Figs.
2(a) and 2(b) can be interpreted in terms of mode coupling and band theory,
since the strength of interlayer coupling and resonance hybridization can be
quantitatively estimated using the overlap integral of the guided resonance
modes of CSRRs on adjacent metallic layers in a unit cell. The strongest
near-field coupling occurs between adjacent layers at a twist angle of
$\theta=0^{\text{o}}$. Moreover, for a twisted system, the 2-fold circular
symmetry of a CSRR maximizes the overlapping regimes where guided resonance
modes exist; thus, such a system is also advantageous for its large coupling
coefficient, which is crucial to band extension.

\begin{figure}[ptb]
\begin{center}
\includegraphics[
width=6.5cm
]{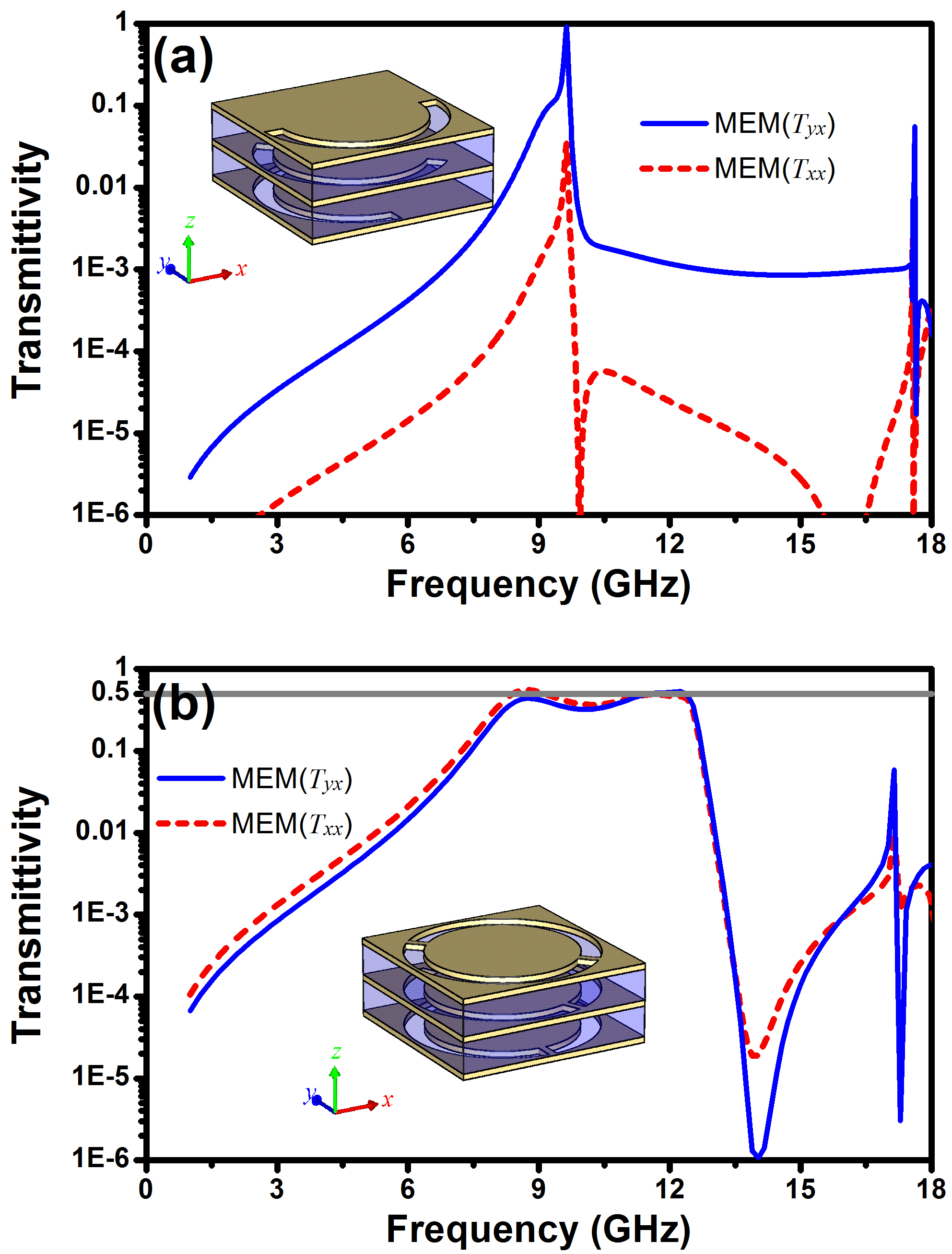}
\end{center}
\caption{Calculated $T_{yx} $ and $T_{xx} $ transmission spectra for the model
systems with three metallic layers of semi-annular rings (a) and CSRRs with
a twist angle of 45\r{ }.(b)}%
\end{figure}

This can be confirmed by control calculations. For instance, let us consider
the periodic systems shown in Figs. 3(c) and 3(d). The two systems are derived
from those of Figs. 3(a) and 3(b), respectively, by removing one semi-annular
ring on each metallic layer in a unit cell. Figures 3(c) and 3(d) show that
the corresponding branches (red lines) become much flatter (i.e., cover a much
narrower frequency range) than those in Figs. 3(a) and 3(b). The lowest dark
band in Fig. 3(a) and the two lowest dark bands in Fig. 3(b) also disappear as
the mirror symmetry of the CSRRs is broken. Figure 4(a) shows the $T_{xx} $
and $T_{yx} $ spectra for the model system shown in Fig. 3(d). Perfect
polarization transformation happens only at a single frequency. The results
further indicate that the circular symmetry of CSRRs, which maximizes the
overlapping aperture regimes, is very helpful for enhancing the strength of
interlayer coupling as well as the BPT functionality.

Furthermore, we emphasize that the 2-fold mirror symmetry of the CSRRs also
ensures that the allowed state propagating through each metallic layer has a
single linear polarization along the orientation of the corresponding CSRRs.
Our CSRR system can rotate the linear polarization of an incident wave in an
arbitrary direction prescribed by the angle $\theta$. Figure 4(b) shows the
transmission spectrum of another sample with a twist angle $\theta
=45^{\text{o}}$. The polarization of the incident wave is changed by
$45^{\text{o}}$ in a frequency range of 8.5$\sim$12.5 GHz.

In summary, an ultrathin multilayer stacked system containing twisted CSRRs is
proposed for manipulating the linear polarization of EM waves and providing
asymmetric transmission in a wide frequency range. The polarization rotation
is rigorously controlled by the twist angle of the CSRRs at opposite ends of
the system. Calculations and experiments both demonstrate that a high
conversion efficiency is obtained in a broad frequency range thanks to the
excitation and hybridization of the guided resonance modes of CSRRs having
circular symmetry. The obtained BPT functionality can be generalized to
terahertz, infrared, and visible frequencies for myriad photonic applications.

This work was supported by NSFC (No. 10974144, 60678046), CNKBRSF (Grant No.
2011CB922001), NCET
(07-0621), STCSM and SHEDF (No. 06SG24).

\end{document}